\documentclass[conference]{IEEEtran}
\IEEEoverridecommandlockouts
\usepackage{cite}
\usepackage{amsmath,amssymb,amsfonts}
\usepackage{algorithmic}
\usepackage{graphicx}
\usepackage{textcomp}
\usepackage{xcolor}
\usepackage{url}
\usepackage{array}
\usepackage{enumitem}
\usepackage{listings}
\usepackage{tcolorbox}
\tcbuselibrary{skins,breakable}

\newcommand{\promptfontsize}{\footnotesize}

\newcommand{\placeholder}[1]{%
  \textcolor{blue!70!black}{\texttt{\{#1\}}}%
}

\newtcolorbox{promptbox}[1][]{
  breakable,
  enhanced,
  colback=gray!5,
  colframe=gray!40,
  fontupper=\ttfamily\promptfontsize,   
  fonttitle=\bfseries\promptfontsize,
  title=#1,
  left=1em, right=1em, top=1em, bottom=1em,
  sharp corners
}

\begin{document}

\title{LLM-Empowered Event-Chain Driven Code Generation for ADAS in SDV systems \\
\thanks{
This research was funded by the Federal Ministry of Research, Technology and Space of Germany as part of the CeCaS project, FKZ: 16ME0800K.
}
}

\author{
\IEEEauthorblockN{
    Nenad Petrovic\IEEEauthorrefmark{1},
    Norbert Kroth\IEEEauthorrefmark{2},
    Axel Torschmied\IEEEauthorrefmark{2},
    Yinglei Song\IEEEauthorrefmark{1},
    Fengjunjie Pan\IEEEauthorrefmark{1},
    Vahid Zolfaghari\IEEEauthorrefmark{1}, \\
    Nils Purschke\IEEEauthorrefmark{1},
    Sven Kirchner\IEEEauthorrefmark{1},
    Chengdong Wu\IEEEauthorrefmark{1},
    Andre Schamschurko\IEEEauthorrefmark{1},
    Yi Zhang\IEEEauthorrefmark{1},
    Alois Knoll\IEEEauthorrefmark{1}
}

\IEEEauthorblockA{\IEEEauthorrefmark{1}
    \textit{Chair of Robotics, Artificial Intelligence and Real-Time Systems} \\
    Technical University of Munich, Munich, Germany \\
    Email: \{nenad.petrovic, yinglei.song, f.pan, v.zolfaghari,
    nils.purschke, sven.kirchner, chengdong.wu, \\
    andre.schamschurko, yi1228.zhang, k\}@tum.de
}

\IEEEauthorblockA{\IEEEauthorrefmark{2}
    CARIAD SE, Wolfsburg, Germany \\
    Email: \{norbert.kroth, axel.torschmied\}@cariad.technology
}

}

\maketitle

\begin{abstract}
This paper presents an event-chain–driven, LLM-empowered workflow for generating validated, event-driven automotive code from natural-language requirements. A Retrieval-Augmented Generation (RAG) layer retrieves relevant signals from large and evolving Vehicle Signal Specification (VSS) catalogs as code generation prompt context, reducing hallucinations and ensuring architectural correctness. Retrieved signals are mapped and validated before being transformed into event chains that encode causal and timing constraints. These event chains guide and constrain LLM-based code synthesis, ensuring behavioral consistency and real-time feasibility. Based on our initial findings from the emergency braking case study, with the proposed approach, we managed to achieve valid signal usage and consistent code generation without LLM retraining.
\end{abstract}


%%%%%%%%%%%%%%%%%%%%%%%%%%%%%
\section{Introduction}

Modern automotive software development faces increasing pressure to support rapid prototyping, iterative refinement, and safe deployment of complex, event-driven functionalities. Traditional development workflows rely on manually created specifications, handcrafted state machines, and domain-specific modeling tools. While these methods provide control and traceability, they also require significant expert effort and are difficult to scale as the number of functions, sensors, and interactions grows \cite{petrovic2025survey}. At the same time, Large Language Models (LLMs) have emerged as powerful tools for translating unstructured requirements into executable artifacts, yet they lack intrinsic guarantees regarding correctness, signal validity, and adherence to timing constraints. Directly generating code from natural-language descriptions using an LLM often leads to hallucinated signal names, inconsistencies with system architecture, or logic that violates real-time execution guarantees.

To bridge this gap, we propose an event-chain–based, LLM-empowered code generation workflow that integrates LLM reasoning with formalized behavioral models. Event chains are widely used in automotive timing analysis and end-to-end latency modeling -- they provide an explicit representation of causal and temporal relationships between system stimuli, internal processing steps, and actuator effects \cite{inc2021} \cite{inc2022}. By embedding event chains within the code-generation pipeline, we can enforce structure, ensure compatibility with the underlying Vehicle Signal Specification (VSS), and validate temporal and logical consistency before any code is produced.

The synergy between event-chain models and LLM-driven generation is central to our approach. Event chains act as behavioral anchors that constrain the LLM’s output: they define allowed signal flows, permissible causal relationships, and timing budgets. The LLM, in turn, contributes flexible interpretation capabilities, mapping natural-language requirements to VSS signals, extracting intent, and synthesizing code templates that conform to the validated event-chain semantics. This combination enables the generation of code that is not only syntactically correct but also behaviorally valid with respect to the system architecture and execution model.

To further reduce hallucinations and ensure fidelity to the authoritative catalog of vehicle signals, we incorporate a Retrieval-Augmented Generation (RAG) layer between the requirement and event-chain construction stages. RAG ensures that the LLM reasons strictly over valid, up-to-date VSS entries, even as the catalog evolves across vehicle generations. Together, RAG, event-chain modeling, and LLM synthesis form a closed-loop workflow that maintains correctness from requirements to deployment.

Ultimately, the proposed pipeline provides a scalable method for transforming natural-language automotive function descriptions into validated, event-driven code. By combining the interpretability and creativity of LLMs with the rigor and structure of event-chain models, the workflow supports rapid iteration while preserving the safety, timing, and architectural constraints required in modern automotive systems. As proof-of-concept for the proposed approach, we show a simplified emergency braking case study where the generated code is executed by a modified ID. Buzz vehicle mounted on the testbench, where hazard lights are activated when a person is detected using the camera.

\section{Background and Related Works}
\subsection{GenAI for SDV Code Generation}
The work presented in this paper builds upon several of our previous works. In \cite{petrovic2025genai}, we presented an end-to-end LLM-driven workflow for ADAS code generation targeting a simulation environment. Synergy with formal methods was used for validation of topological aspects in vehicular system - underlying components, their properties and relations. This way, it is possible to determine whether the system configuration is acceptable for intended ADAS capabilities. On the other side, in \cite{pan2025}, we incorporate complementary elements related to behavioral aspects of vehicles in the context of ADAS. Here, we make use of a simple event chain-based representation for validation of basic constraints, such as order of events. The target code was aiming CARLA open source simulation environment for ADAS capabilities, incorporating ROS support for integration of generated modules with rest of the system, thanks to message publish/subscribe mechanism. 
Furthermore, in \cite{zyberaj2025genai}, we presented GenAI-empowered workflow considering VSS signals and Gherkin language for test scenario representation and validation before Python code generation aiming digital.auto SDV testing and simulation platform. Additionally, our work from \cite{pavel2025hallucination} is focused on tackling hallucinations when it comes to automotive platform code generation. In that work, relevant prompting strategies are first identified, while their practical usability in various cases is evaluated in the later part. On the other side in \cite{petrovic2025multimodal}, we explore the adoption of Vision Language Models (VLMs) for extracting requirements and relevant information from automotive-related diagrams. Moreover, in \cite{zolfaghari2024} we presented Retrieval Augmented Generation (RAG) methodology based on Retrieve and Re-Rank for compliance checking with respect to large textual documents (standards and UN regulations) in the context of automotive software and hardware design. 
In this paper, we want to integrate main findings of those works into end-to-end code generation framework aiming at the target automotive platform for ADAS testbench \cite{wu2025viltum}, starting from vehicle behavior description as freeform text. Event chain-based representation will be used for validation purposes, as well as agent memory within the iterative process of user feedback and refinement of vehicle behavior description.

\subsection{SDV Exploration Environment}
\begin{figure}[t]
    \centering
    \includegraphics[width=\columnwidth]{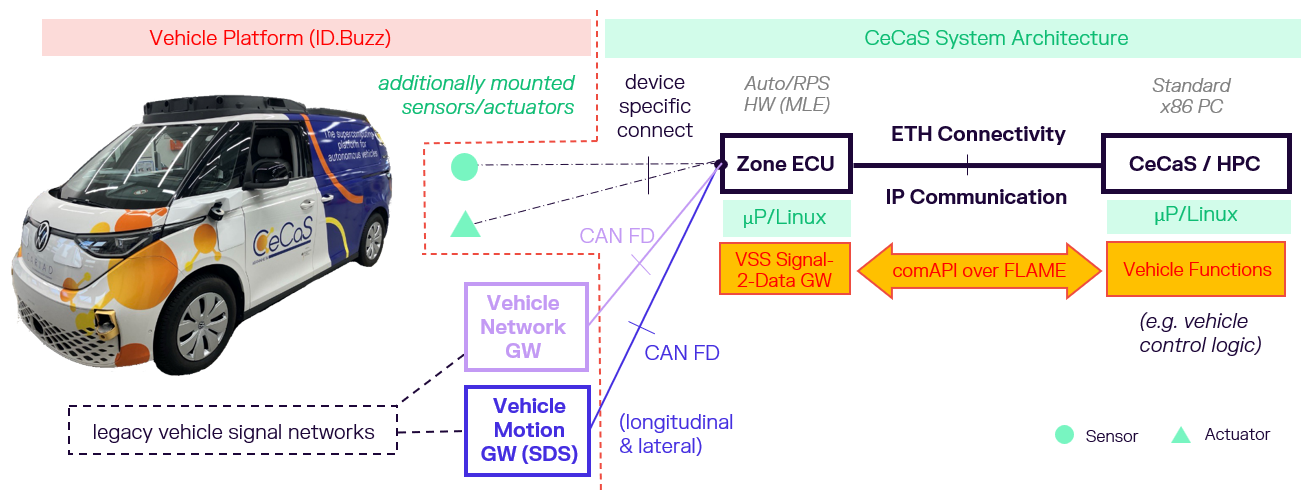}
    \caption{Structure of the automotive exploration environment}
    \label{fig:cariad1}
\end{figure}

The automotive exploration environment to evaluate the generated code of the LLM-driven workflow is based on a zonal architecture for the SDV, which has already been described in \cite{cariad1}. The target system used for the test has been specifically built on a modified vehicle platform (ID.Buzz) as shown in Fig. \ref{fig:cariad1}. The IT-system that was added to the car includes a zone controller (ZoneECU), a high-speed Ethernet backbone, IP-based communication via the FLAME protocol, and a standard PC, representing a central high-performance compute platform (HPC). The “Vehicle Motion Gateway” enables setting up a redundant self-driving system (SDS) in order to move the car (longitudinal, lateral), whereas the “Vehicle Network Gateway” allows to interact with other vehicle sub-systems (e.g., body functions like hazard-lights, etc.). Both special-purpose ECUs offer a dedicated CAN FD interface to access and manipulate corresponding vehicle signals via a data-centric vehicle interface (Signal-2-Data Gateway). The Gateway is a server instance that acts as an interface to the legacy vehicle, which converts VSS-based data structures into CAN signals (and vice versa).

\subsection{Vehicle Communication Middleware}
\begin{figure}[t]
    \centering
    \includegraphics[width=\columnwidth]{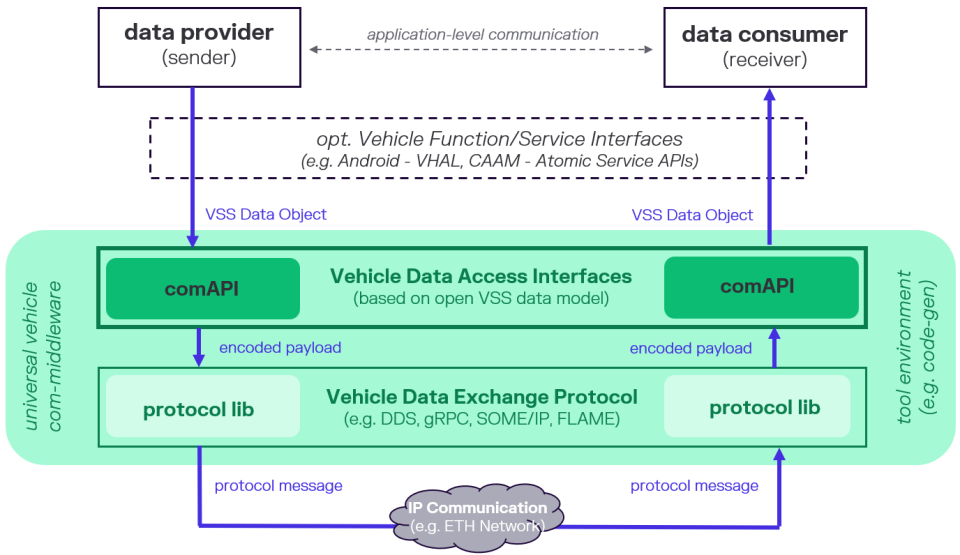}
    \caption{Structure of the universal vehicle communication middleware.}
    \label{fig:cariad2}
\end{figure}

In \cite{cariad2}, it was presented that the COVESA VSS specification (incl. semantic vehicle models) was used to create an abstract “vehicle data space” in the CeCaS software stack. The universal vehicle communication middleware separates data access from the transport layer (see Fig. \ref{fig:cariad2}). This provides stability for the application interface and flexibility on the protocol layer (multiple protocol binding). The VSS-based data objects are accessed at the application layer by provider and consumer entities via an abstract, protocol-agnostic, data-centric interface (comAPI). In the CeCaS exploration environment, the payload is transported over the FLAME protocol between distributed nodes (ECUs) in the system. 

By using the proposed LMM-based, event-chain driven approach, we’ve practically evaluated how the proposed workflow can be used in combination with the VSS-based comAPI to auto-generate code for automotive functions.

\section{Implementation overview}

\subsection{Workflow description}
\begin{figure*}[htbp]
\centering
\includegraphics[width=\textwidth]{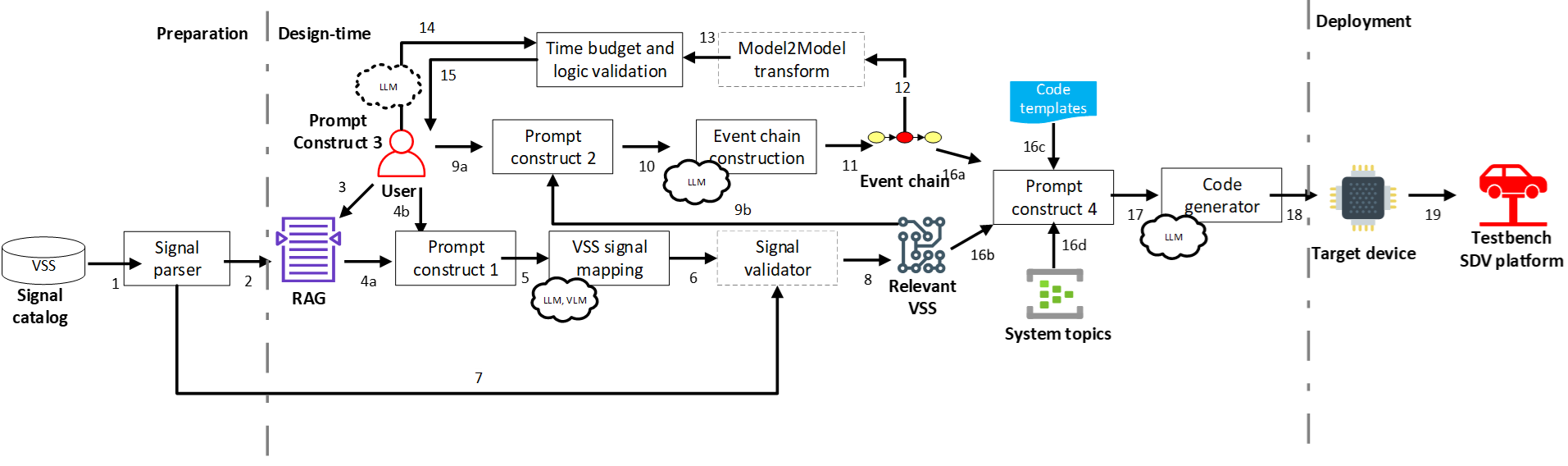}
\caption{Workflow of event chain-driven LLM-empowered ADAS code generator.}
\label{event_chain_workflow}
\end{figure*}

The presented workflow defines an automated pipeline for LLM-empowered generation of event-driven code for ADAS capabilities of SDV systems by combining VSS-based signal mapping and model-validated event-chain synthesis. The process is organized into three major stages -- Preparation, Design-time, and Deployment, while each step within these stages corresponds to a numbered element in the accompanying diagram, as shown in Fig. \ref{event_chain_workflow}.

\textit{Preparation (step 1)}: The workflow begins with a curated Vehicle Signal Specification (VSS) catalog containing all available vehicle signals, which can be quite numerous. Signal Parser (step 1) processes the VSS catalog and extracts only the relevant information for code generation. For each of the signals, apart from name, we also extract the information about signal type (sensor/actuator), as well as relevant getter and setter methods in comma-separated format \textit{(signal, type, getters/setters)}, such as in case of hazard lights:
\texttt{Vehicle.Body.Lights.Hazard, actuator, set\_is\_signaling(bool value)}

\textit{RAG-based context construction (step 2)}: Between the Preparation and Design-time stages, we employ Retrieval-Augmented Generation (RAG) in step 2 to construct contextualized, VSS-aware inputs for the downstream LLM components. RAG is essential in this setting for several reasons. First, VSS catalogs can contain thousands of signals, attributes, and hierarchical paths, which typically exceed the LLM’s context window. Second, VSS evolves over time; new signals are added, naming conventions change, and domain-specific terminology often does not appear in the model’s pre-training corpus. Relying solely on the LLM’s internal knowledge would therefore increase the risk of hallucinations, incorrect signal names, outdated terminology, and synonym confusion. To address this, we adopt a retrieve-and-re-rank pipeline similar to \cite{zolfaghari2024}. The hierarchical VSS JSON structure is first flattened into a one-entry-per-line knowledge base, ensuring that each signal definition becomes an independent retrievable unit. Both the flattened entries and the user-provided scenario description are embedded into dense vector representations. We use a SentenceTransformer model (all-MiniLM-L6-v2 \cite{allMiniLM2024}) to compute these embeddings, as it captures semantic similarity between phrases such that related concepts -- e.g., “acceleration,” “pedal position,” “torque request” -- lie close together in the embedding space. During retrieval, the system selects the top-k most similar VSS entries to the given scenario query. These candidates are then passed to a re-ranking step, which uses either a cross-encoder model (ms-marco-MiniLM-L-6-v2 \cite{msmarcoMiniLM2025}) or lightweight domain-specific rules. The cross-encoder directly compares each candidate with the scenario text and assigns a relevance score, enabling more precise filtering. Rule-based signals (e.g., “accelerating” implying speed, pedal, and torque-related signals) further improve domain-aware prioritization. The highest-ranked entries are subsequently chunked to fit within the LLM input limits. The LLM is queried multiple times -- once for each chunk -- together with the scenario description. The model outputs are then normalized, merged, and de-duplicated, producing a consolidated list of VSS signals relevant to the scenario. Because the LLM is constrained to generate from retrieved, valid entries, the approach significantly improves precision, reduces hallucinations, and ensures that the output remains consistent with the authoritative VSS catalog. Overall, this RAG-based context construction enables scalable, catalog-aware reasoning without retraining the LLM, while robustly handling large, evolving signal definitions within the workflow.

\textit{User intent interpretation (steps 3-5)}: The user provides a natural-language description of the intended behavior or scenario at the beginning of the design phase. The RAG system retrieves relevant signal definitions in order to construct an appropriate context for code generation. A Prompt Construction 1 module synthesizes the first LLM prompt. The following prompt is constructed based on the user-provided scenario and context retrieved using RAG. Handling of visual inputs is built upon our previous work from \cite{petrovic2025multimodal}.

\begin{promptbox}[Prompt Construction 1 -- VSS Signal Selection]
Based on the following ADAS scenario: \placeholder{scenario/diagram}, select all relevant Vehicle Signal Specification (VSS) signals. \\

The full list of relevant VSS signals is: \placeholder{VSS candidate signals}. The subset of signals from the list that are relevant should be returned as a comma-separated list (without comments and explanations).
\end{promptbox}

\textit{Signal mapping and validation (steps 6-9)}: Previously constructed Prompt Construct 1 is executed against the target model. LLM or VLM maps the extracted intent to corresponding VSS signals from the previously prepared catalog, producing a candidate signal list. Those signals are expected to appear within the generated code in the final step. Optionally, Signal Validator checks these mappings for correctness, ensuring that the selected signals are presented in the original full VSS catalog. After that, validated signals are ready to be further used as inputs for code generation.

\textit{Event-chain construction (Steps 10–15)}: LLM generates a structured event-chain specification starting from textual user-provided scenario, describing causal, temporal, and conditional relationships. For this purpose, we use PlantUML format for UML activity diagram-alike representation. This choice was made due to simplicity of notation and possibility of using the existing visualization tools in order to show the intermediate result to the user. UML activity diagram is enriched with notes that include additional useful information for event chain validation, such as time budget (in ms), expected input, generated output, as well as input and output formats.
For this reason, the prompt is constructed as follows:

\begin{promptbox}[Prompt Construction 2 -- Event chain generation based on user scenario.]
You are updating PlantUml activity diagram about automotive event chain without comments and without explanations given as \placeholder{current-event-chain}, with respect to user scenario: \placeholder{scenario}. \\

For each of the steps in the event chain, the following parameters are considered as notes: time\_budget, input, input\_format, output, output\_format. Event chain should be as simple as possible, it is assumed that messages are received from simulator and executed on vehicle. Only consider steps affecting the decision from perspective of vehicle.
\end{promptbox}

On the other side, model transformation step converts the initially constructed event chain into a formal internal representation suitable for validation, which is simple JSON-based model in our case. The Model2Model Transform ensures compatibility with downstream timing and execution models, potentially providing support to other external validation and event chain analysis tools as well. After that, the event-chain model undergoes timing-budget validation, analyzing latency and execution order. Logic validation ensures semantic correctness, verifying that the event chain is executable given the available signals and constraints. For purpose of validation, we also make use of LLM in order to transform freeform textual description to validation rules that will be checked against the JSON form of the event chain. Two types of simple event chain validation rules are supported: \textit{order - before/after} and \textit{time budget} rules. The first type of rules is used to check the order of events within the chain. If a rule is defined as \textit{$e_1 \ \mathbf{before} \ e_2$}, 
it is expected that event $e_1$ happens before $e_2$, and vice versa in the case of the \textit{after} variant. On the other hand, time-budget rules have the form 
\textit{execution-time($e_1) \le \text{time-budget-limit}$}, which means that the estimated execution time for a particular event $e_1$ should not exceed the given limit. For these time-related constraints, the assumed measurement is done in milliseconds (ms). In what follows, Listing 3. corresponding to step 14 is given.

\begin{promptbox}[Prompt Construction 3 -- Event chain validation rules generation]
Based on textual input \placeholder{constraint rule}, generate a shell script that invokes the Python script event\_chain\_validator.py, which always requires: --json event\_chain.json \\

Rules can be defined as: \\
1) --rule "order:action1 before action2" \\
2) --rule "order:action1 after action2" \\
3) --rule "time:action <= [time\_budget]" (time\_budget is integer) \\

For each identified rule, the shell script should make another call to event\_chain\_validator.py with the corresponding --rule argument. \\

Create the simplest possible minimal shell script that calls event\_chain\_validator.py with the appropriate arguments.
\end{promptbox}

\textit{Code Generation (steps 16-18)}: The validated event chain and relevant VSS information are passed to a third prompt construction stage. The main output from these steps is decision-making code that aims to be executed on the specific vehicular module in order to implement the desired ADAS capability. Code templates, as well as language and platform patterns for target device (ECU, zone controllers, HPC) are incorporated to ensure that all the required dependencies are considered, the corresponding structure of code is satisfied, reducing possible hallucinations this way. On the other side, available messages within the rest of the system are also provided as input, in order to successfully integrate the generated software module with the rest of the SDV system. In our experiment, we used MQTT protocol for message exchange, so the prompt input will therefore also include a list of candidate topics which can be leveraged in order to retrieve the required information for ADAS-related decision-making (such as whether LIDAR or camera detected a pedestrian). Listing 4 shows the prompt used for code generation. As it can be seen, in order to ensure the correctness of VSS signal handling, we make use of In-Context Learning technique, by incorporating task solution example covering this aspect as part of the code generation prompt.

\begin{promptbox}[Prompt Construction 4 -- Automotive C++ Testbench Generation]
You are generating C++ code for an automotive testbench based on a vehicle behavior description. All generated code concerns decision-making logic from the vehicle's perspective. Sensing is assumed to occur by receiving MQTT messages from the list:
\placeholder{system-topics} \\

Actuation on the real vehicle is performed by selecting VSS signals from: \placeholder{Extracted VSS signals} \\

For each selected VSS signal, use the following mapping pattern, like given for acceleration: \\

1) In header: \\
   \#include <capnproto/types/vehicle/acceleration.h> \\

2) In void call\_runner(): \\
   auto set\_acceleration = create\_caller<types::vehicle::Acceleration>(); \\

3) Declare actuator value: \\
   types::vehicle::Acceleration acceleration\_value\{\}; \\

4) Modify actuator value: \\
   acceleration\_value.set\_longitudinal(
       acceleration\_value.longitudinal() + 0.2
   );
   ret = set\_acceleration.call(acceleration\_value, \&result\_id); \\

The final script must remain minimal: no keyboard interaction, only MQTT message reception, decision logic, and calls to VSS handlers. Print messages should be preserved for debugging when a message is 
received or an actuator is triggered. \\

Use the extended example as reference for structure and dependencies: \placeholder{code-template}.
\end{promptbox}

\textit{Deployment Phase (step 19)}: The generated code is deployed to the target embedded device (e.g., automotive ECU or zone controller) - either directly or over the air. In our case, we use CARIAD's zone controller running commAPI middleware which maps VSS signals to CAN messages in order to control the vehicle. The zone controller is accessed over the air via SSH, the generated code transferred and software re-built.

\begin{figure}[t]
    \centering
    \includegraphics[width=\columnwidth]{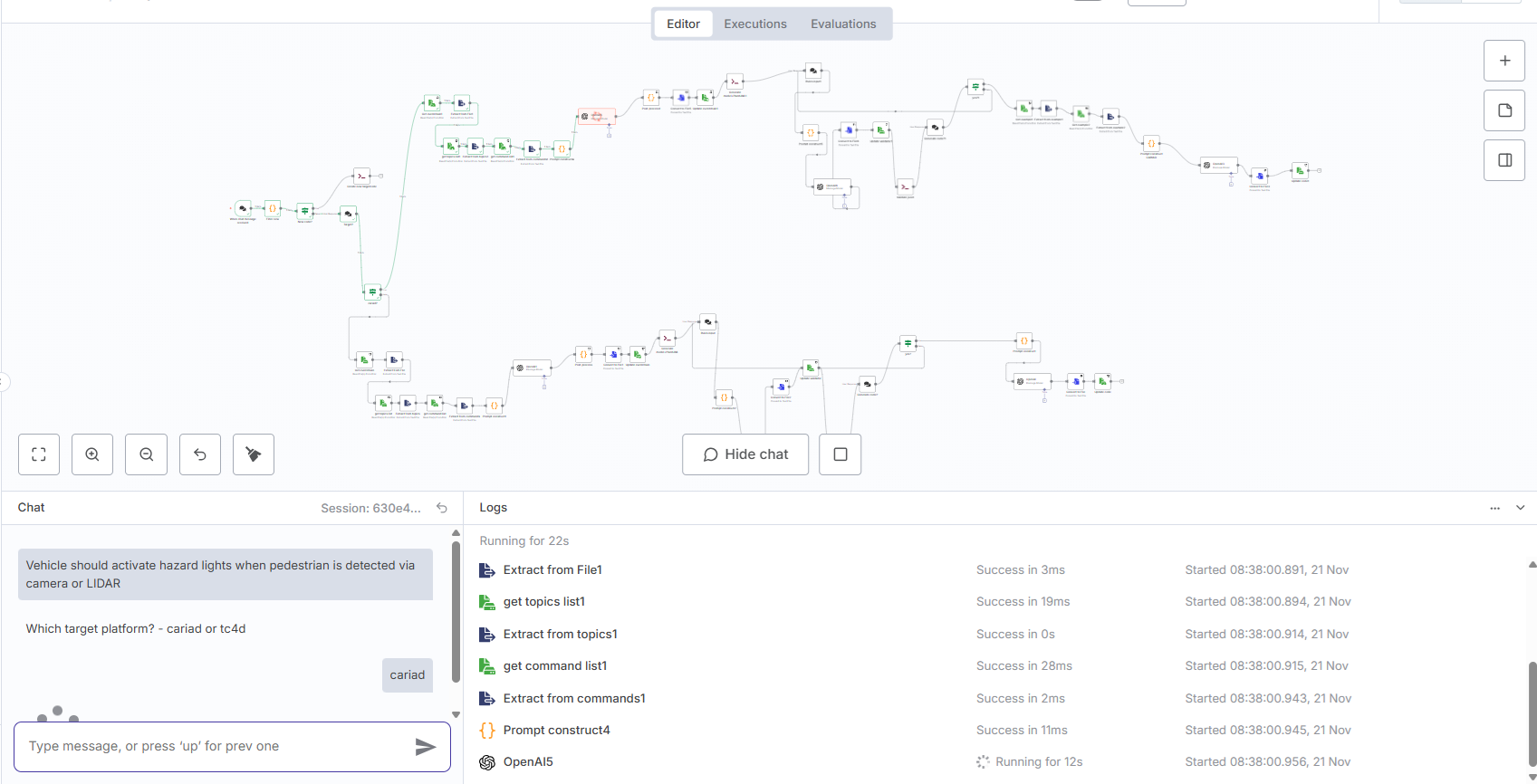}
    \caption{Integrated code generation toolchain based on n8n.}
    \label{fig:n8n_wf}
\end{figure}

The final system is executed and tested within a testbench Software-Defined Vehicle (SDV) platform, enabling integration testing, simulation, and iterative refinement.
Screenshot of the integrated toolchain within the n8n workflow tool is given in Fig. \ref{fig:n8n_wf}.

\section{Case Study Overview}
\begin{figure*}[t]
    \centering
    \includegraphics[width=\textwidth]{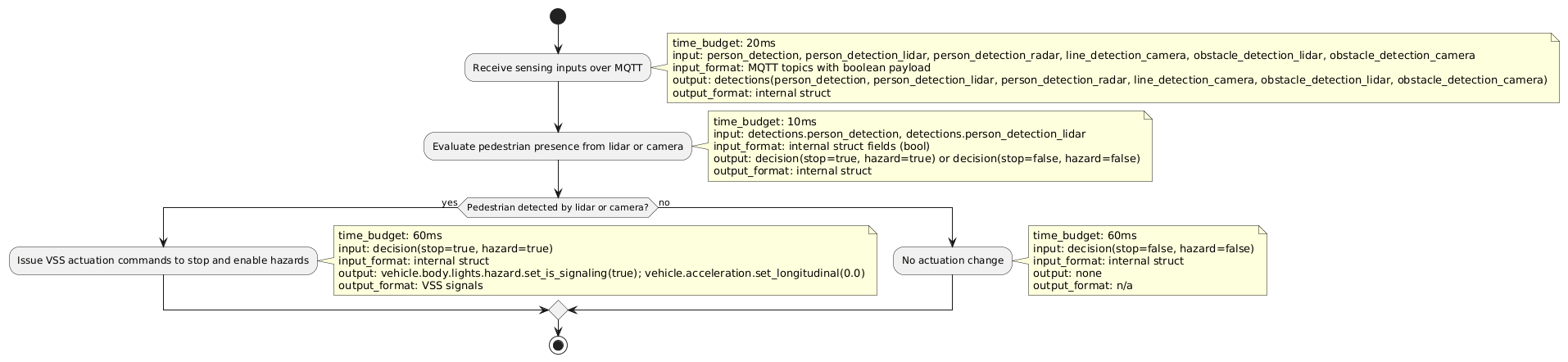}
    \caption{Generated event chain for case study example, rendered as PlantUML activity diagram.}
    \label{fig:event_chain_example}
\end{figure*}

In order to show proof-of-concept, we define the following scenario that is considered as input of the code generation toolchain:
\textit{"Vehicle should activate hazard lights when camera or LIDAR detects a pedestrian."}
Generated event chain is depicted in Fig. \ref{fig:event_chain_example}.

The following validation rules were checked against the generated event chain: 1) camera before hazard; 2) hazard after receiving; 3) $hazard \leq 90$ - reaction to pedestrian detection should be performed within 90ms.

Full VSS catalog, in our case, contained 64 entries to select from. On the other side, there were also several MQTT topics considered in order to enable integration of the generated decision logic with the rest of the system: camera-front-detect, camera-back-detect, and lidar-detect.

For camera-based detection, we run a service in a Docker container based on YOLOv11~\cite{yolo}, which generates an MQTT message with the value "1" on the camera-related topic if the person is detected. Therefore, the generated decision logic code is listening to this relevant topic and executing activation of hazard lights via VSS signal mapping, thanks to commAPI.

%YINGLEI SHOULD ADD SOME TEXT ABOUT THAT OR IMAGE OR ANYTHING ELSE HERE ABOUT DETECTION

\begin{figure}[t]
    \centering
    \includegraphics[width=\columnwidth]{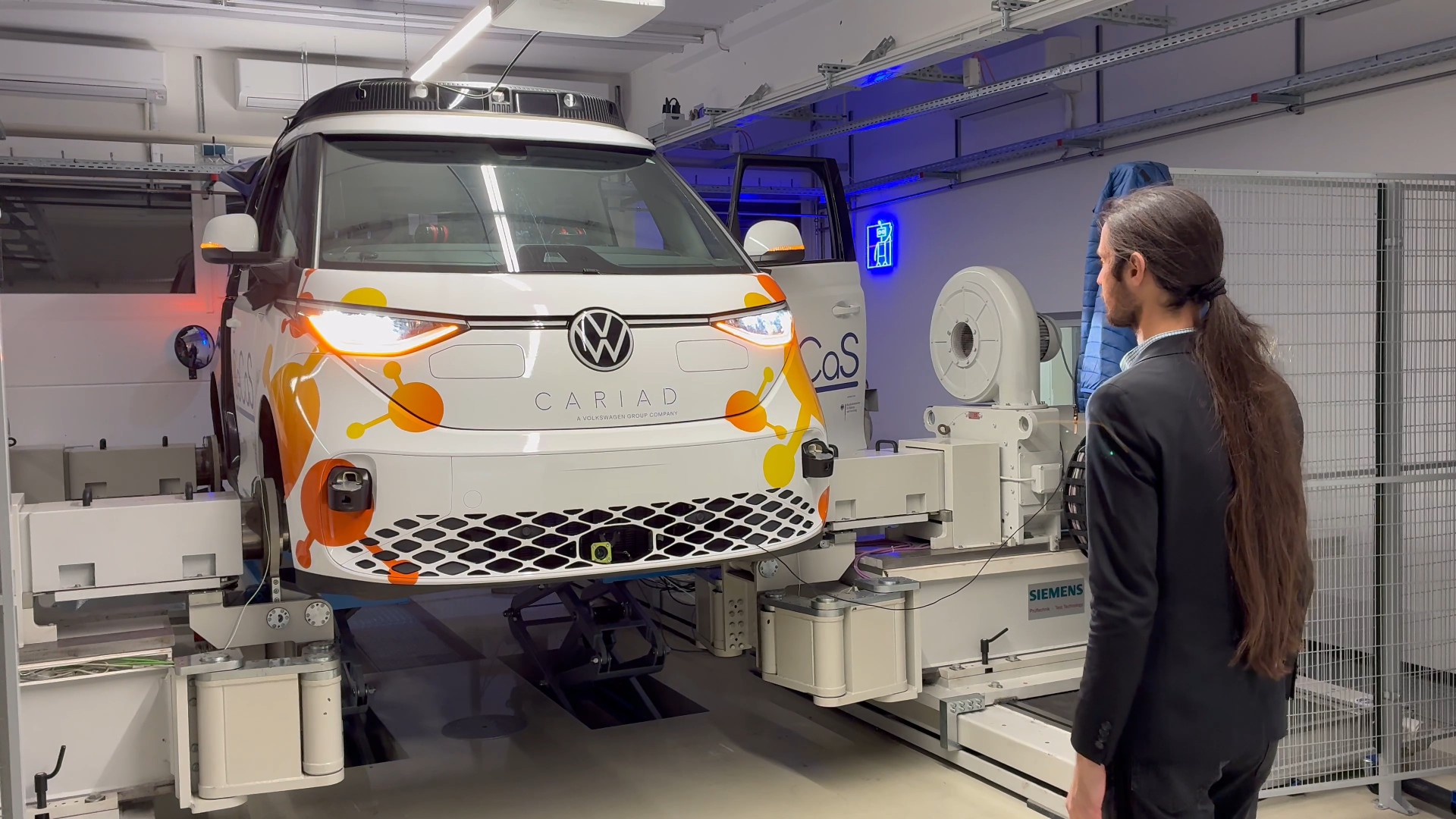}
    \caption{Case study illustration: Hazard lights activation when a pedestrian is detected.}
    \label{fig:case_study}
\end{figure}

\begin{figure}[t]
    \centering
    \includegraphics[width=\columnwidth]{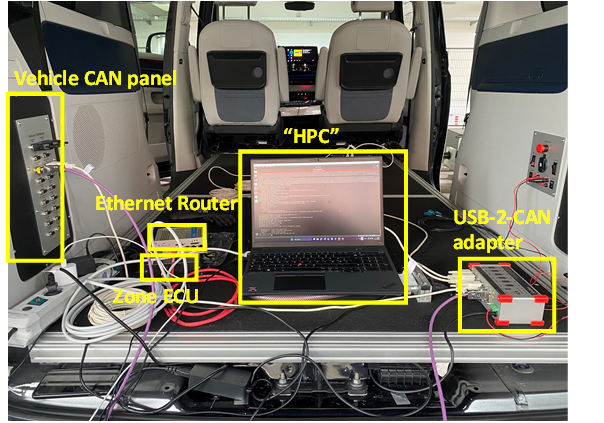}
    \caption{In-vehicle hardware running the code.}
    \label{fig:hw_setup}
\end{figure}

The outcome of the code running on target platform for the case study is shown in Fig. \ref{fig:case_study}. Hardware configuration inside the vehicle, with highlighted key components running the generated code inside the vehicle, is shown in Fig. \ref{fig:hw_setup}.

\section{Conclusion}
In this paper, we show a proof-of-concept implementation of code generation methodology relying on LLMs, aiming at a physical SDV platform.
According to the initial findings, the main aspect contributing to practical usability of LLM-based toolchains in such scenarios is breaking down the complex problem into simple, verifiable steps based on formally grounded methodologies. In our case, we aim to ensure design-time correctness of the generated code as much as possible, making use of model-driven representation based on event chain concept.
%%%%%%%%%%%%%%%%%%%%%%%%%%%%%

\vspace{12pt}

\end{document}